\documentclass[11pt,draftcls,onecolumn]{IEEEtran}              

\usepackage{graphicx}

\usepackage{cite}
\usepackage{amssymb}
\usepackage{amsthm}

\newtheorem{theorem}{Theorem}

\newtheorem{proposition}{Proposition}

\usepackage[cmex10]{amsmath}
\begin{document}

\title{On Fast Algorithm for Computing Even-Length~DCT}

\author{Yuriy~A.~Reznik
\thanks{The author is with Qualcomm~Inc., San~Diego, CA, 92121 USA;  \mbox{e-mail}:~yreznik@qualcomm.com.}}


\date{December 29, 2009}

\maketitle

\begin{abstract}
We study recursive algorithm for computing DCT of lengths $N=q\, 2^m$ ($m,q \in \mathbb{N}$, $q$ is odd) due to C.\,W.~Kok~\cite{kok}. We~show that this algorithm has the same multiplicative complexity as theoretically achievable by the prime factor decomposition, when $m \leqslant 2$. We~also show that C.\,W.~Kok's factorization allows a simple conversion to a scaled form. We analyze complexity of such a scaled factorization, and show that for some lengths it achieves lower multiplicative complexity than one of known prime factor-based scaled transforms~\cite{FeigLinzer}.
\end{abstract}

\begin{IEEEkeywords}
Discrete cosine transform, DCT, scaled transform, factorization, multiplicative complexity.
\end{IEEEkeywords}

\IEEEpeerreviewmaketitle

\section{Introduction}
\IEEEPARstart{T}{he} {\em discrete cosine transform\/} (DCT)~\cite{AhmedNatarajanRao,RaoYip,BritanakYipRao} is a fundamental and frequently used operation in modern digital signal processing. It~finds applications in data compression, filter design, image recognition,~etc. {\em Scaled\/} DCT is a modified version of this transform, allowing the output to be scaled in way that simplifies its computation~\cite{FeigWinograd}. Scaled DCTs are particularly popular in data compression, where scaling of DCT output can usually be done jointly with quantization, therefore reducing the complexity of the entire algorithm~\mbox{\cite{AAN,Feig}}.


Since its discovery in early 1970s, DCT has been a subject of extensive research, focusing, in part, on the design of fast algorithms for its computation~\mbox{\cite{RaoYip,BritanakYipRao}}. The class of DCT of type II (DCT-II) with dyadic lengths $N=2^m$, $m\in{\mathbb N}$ has been studied particularly well. Both theoretical complexity estimates \cite{Duhamel,HeidemanBurrus,FeigWinograd} and a number of efficient algorithms for their construction have been derived~\cite{
ChenSmithFralick,Wang83,BGLee,LLM,FeigWinogradSP}.
The construction of scaled DCT-II of dyadic lengths has also been studied~\cite{AAN,Feig,FeigWinograd}.
Scaled factorizations of Y. Arai, T.Agui and M. Nakajima~\cite{AAN} and E.~Feig and S.~Winograd~\cite{FeigWinograd} are among best-known algorithms from this class.
The construction of odd-length transforms has been studied by M.~Heideman~\cite{Heideman}, and S.~Chan and K.~Ho~\cite{ChanHo93}, uncovering, in part, an elegant connection between real valued DFT and DCT-II of same lengths.
The construction of DCT of composite sizes, such as $N=p\,q$, where $p$ and $q$ are co-prime, was studied by P.~Yang and M.~Narasimha~\cite{YangNarasimha}, B.G. Lee~\cite{BGLee89},
and others, resulting in the development of the {\em prime-factor decomposition\/} of the DCT-II. E.~Feig and E.~Linzer have further extended the prime-factor technique for computing scaled transforms~\cite{FeigLinzer}.
The construction of DCT of even (but not dyadic) lengths has been addressed by a variety of techniques, ranging from prime-factor decompositions~\cite{YangNarasimha,BGLee89} to generalizations of the radix-$2$ DCT algorithm~\cite{kok}. Reference~\cite{kok} contains comparison of several such approaches.

In this correspondence we take another look at the recursive algorithm for computing of DCT of lengths $N=q\, 2^m$, $m,q \in \mathbb{N}$,
proposed by C.\,W.\,Kok~\cite{kok}.
We offer an alternative matrix formulation of this algorithm, its detailed complexity analysis, and a modification allowing to compute a scaled DCT. We~show, that C.\,W.\,Kok's algorithm achieves the same multiplicative complexity as one theoretically attainable by the prime factor decomposition when $m\leqslant 2$. We also show that for some lengths our proposed scaled version of C.\,W.\,Kok's algorithm achieves lower multiplicative complexity than one of known scaled prime factor algorithm-based factorizations~\cite{FeigLinzer}. We accompany our presentation with several examples of scaled factorizations constructed by using described algorithms, and complexity comparison plots that can be of interest to the engineering community.

This correspondence is organized as follows. In~Section~\ref{sec:II}, we introduce notation and survey relevant results. In~Section~\ref{sec:III}, we offer matrix formulation of C.\,W.\,Kok's algorithm and its complexity analysis. An modified (scaled) version of C.\,W.\,Kok's factorization is described in Section~\ref{sec:IV}. Section~\ref{sec:V} contains comparison with prime factor-based implementations. Section~\ref{sec:VI} brings remarks on normalized multiplicative complexity of composite-length transforms. Conclusions are drawn in Section~\ref{sec:VII}.

\section{Notation and Some Basic Facts} \label{sec:II}
By $C^{II}_N$, $C^{III}_N$ and $C^{IV}_N$ we will denote matrices of $N$-point DCT-II, DCT-III, and DCT-IV transforms correspondingly\footnote{For simplicity, we omit normalization factors~\cite{RaoYip}. }
\begin{equation*}
\left.
\begin{array}{rcl}
\left[C^{II}_N\right]_{n,k} & = & \cos\left(\frac{\pi\,(2n+1)\,k }{2N}\right)_{_{\,}} \\
\left[C^{III}_N\right]_{n,k} & = & \cos\left(\frac{\pi\,n\, (2k+1)}{2N}\right)_{_{\,}} \\
\left[C^{IV}_N\right]_{n,k} & = & \cos\left(\frac{\pi\,(2n+1)(k+1)}{4N}\right)
\end{array}
\right\} ~ n,k=0\mathrm{,...,}N\mathrm{-}1.
\end{equation*}
Among these transforms, the DCT of type II (DCT-II) is the one that we will need to compute.

It is well known (see, e.g. \cite{RaoYip}), that the DCT-III is simply an inverse (or transpose) of DCT-II
\begin{equation*}
C^{III}_N = \left(C^{II}_N\right)^T = \left(C^{II}_N\right)^{-1}\,,
\end{equation*}
and that DCT-IV is involutary (self-inverse, self-transpose)
\begin{equation*}
C^{IV}_N = \left(C^{IV}_N\right)^T = \left(C^{IV}_N\right)^{-1}\,. \label{eq:invol}
\end{equation*}

It is also known (cf. S.\,C.\,Chan and K.\/L.\,Ho~\cite{ChanHo}, C.\,W. Kok~\cite{kok}), that \mbox{DCT-IV} and \mbox{DCT-II} are connected as follows
\begin{equation}
C^{IV}_N = R_N C^{II}_N D_N\,, \label{eq:dct4_via_dct2}
\end{equation}
where $R_N$ is a matrix of recursive subtractions
\begin{equation*}
R_N = \left(
\begin{array}{rrrcr}
\tfrac{1}{2} & 0 & 0 & \ldots & 0 \\
-\tfrac{1}{2} & 1 & 0 & \ldots & 0 \\
\tfrac{1}{2} & -1 & 1 & \ldots & 0 \\
\vdots & \vdots & \vdots & \ddots & \vdots \\
-\tfrac{1}{2} & 1 & -1 & \ldots & 1
\end{array}
\right)\,, \label{eq:R_N}
\end{equation*}
and $D_N$ is a diagonal matrix
\begin{equation*}
D_N =
\left(
\begin{array}{cccc}
\!\!2\cos\left(\frac{\pi}{4N}\right)\!\! & & & 0 \\
   & \!\!2\cos\left(\frac{3\pi}{4N}\right)\!\!\!\! & & \\
   & & \ddots & \\
 0 & & & \!\!\!\!2\cos\left(\frac{(2N-1)\pi}{4N}\right)\!\!\!
\end{array}\!
\right). \label{eq:D_N}
\end{equation*}

Further, it is known (cf. W.\,H.\,Chen, et. al.~\cite{ChenSmithFralick}, Z.\,Wang~\cite{Wang83}) that if length of DCT is even, then it can be factored into two half-length transforms
\begin{equation}
C^{II}_N = P_N
\left(
\begin{array}{cc}
C^{II}_{N/2} & 0 \\
0 & C^{IV}_{N/2} J_{N/2}
\end{array}
\right)
B_N, \label{eq:dct2_basic_split}
\end{equation}
where $P_N$ is a permutation matrix, producing reordering
\begin{equation*}
\tilde{x}_i = x_{2i}, ~~ \tilde{x}_{N/2+i} = x_{2i+1}, ~~~i=0,\ldots,N/2-1,
\end{equation*}
$B_N$ is a butterfly
\begin{equation*}
B_N =
\left(
\begin{array}{cc}
I_{N/2} & J_{N/2} \\
J_{N/2} & -I_{N/2}
\end{array}\!
\right), \label{eq:B_N}
\end{equation*}
and where $I_{N/2}$ and $J_{N/2}$ denote $N/2 \times N/2$ identity and order reversal matrices correspondingly.

\section{Recursive DCT-II Computation. Algorithm of C.\,W.\,Kok~\cite{kok}} \label{sec:III}
We note that factorization~(\ref{eq:dct2_basic_split}) is not fully recursive: it uses  \mbox{DCT-II} and \mbox{DCT-IV} of length $N/2$ as building blocks, but subsequent factorization of \mbox{DCT-IV} is not defined. One possible way of closing the recursion is to simply replace \mbox{DCT-IV} with \mbox{DCT-II} in accordance with~(\ref{eq:dct4_via_dct2}). This way we arrive at the following factorization:
\begin{equation}
C_N^{II} = P_N
\left(
\begin{array}{cc}
\!C^{II}_{N/2} & 0 \\
0 & R_{N/2} C^{II}_{N/2} D_{N/2} J_{N/2}
\end{array}
\right)
B_N\,. \label{eq:kok}
\end{equation}
This factorization can be applied recursively, producing a simple algorithm for computing of DCT-II of even lengths, known as C.\,W.\,Kok's algorithm~\cite{kok}.

\begin{figure}[!t]
\centering
\resizebox{!}{2.5in}{\includegraphics{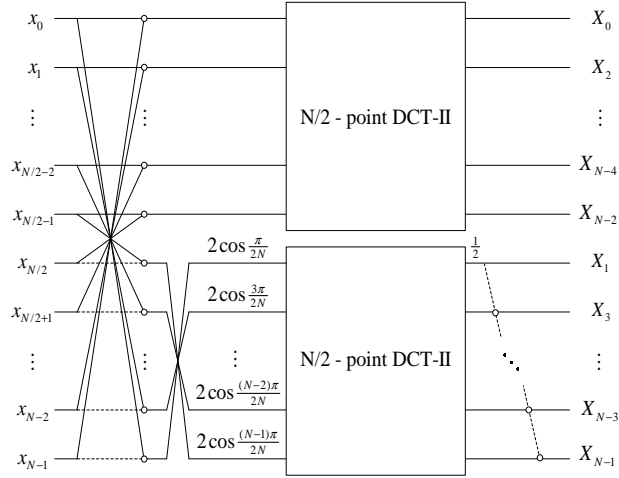}}
\caption{C.\,W.\,Kok's factorization of even-length DCT-II~\cite{kok}.}
\label{fig_1}
\end{figure}

We show the flowgraph of this algorithm in Fig.~\ref{fig_1}. As~customary, dashed lines in the flowgraph denote sign inversions, circles indicate additions, and constants above lines indicate multiplications by the corresponding factors.

Based on Fig.~\ref{fig_1}, it can be observed that the numbers of multiplications~$\mu(N)$, additions and subtractions~$\alpha(N)$, and shifts (multiplications by dyadic factors)~$\sigma(N)$ satisfy
\begin{eqnarray*}
\mu(N)    & = & 2 \mu(N/2) + \tfrac{1}{2} N\,, \\
\alpha(N) & = & 2 \alpha(N/2) + \tfrac{3}{2} N -1\,, \\
\sigma(N) & = & 2 \sigma(N/2) + 1\,.
\end{eqnarray*}

By applying this decomposition recursively $m$-times, we arrive at the following result (cf.~\cite{kok}).
\begin{proposition}[C.\,W.\,Kok, 1997]
The numbers of arithmetic operations ($\mu,\alpha,\sigma$) needed for computing DCT-II of length $N=q\,2^m$ using C.\,W.\,Kok's algorithm, satisfy:
\begin{eqnarray}
\mu(N)    & = & 2^m \mu(q) + \tfrac{m}{2} N \,, \label{eq:mu_kok} \\
\alpha(N) & = & 2^m \alpha(q) + \tfrac{3 m}{2} N - 2^m +1\,, \nonumber \\ 
\sigma(N) & = & 2^m \sigma(q) + 2^m-1\,. \nonumber
\end{eqnarray}
\end{proposition}


\begin{figure}[!t]
\centering
\resizebox{!}{2.5in}{~~\includegraphics{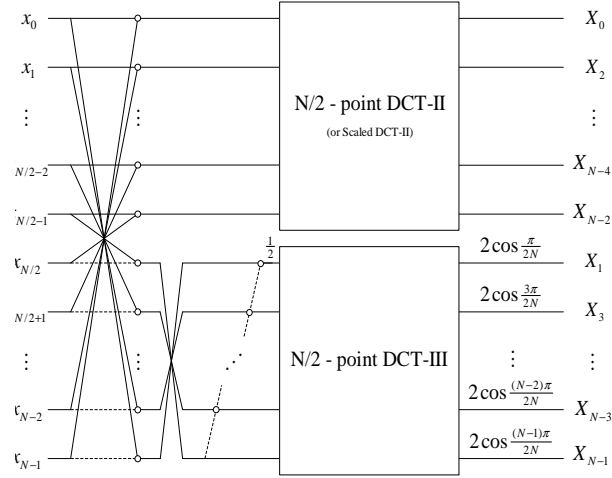}}
\caption{Proposed alternative factorization of even-length DCT-II.}
\label{fig_2}
\end{figure}

\section{Proposed Alternative (Scaled) Factorization} \label{sec:IV}
Consider DCT-II factorization (\ref{eq:dct2_basic_split}) one more time. Since DCT-IV is involutary, we can compute it in a transposed fashion, producing (cf. (\ref{eq:dct4_via_dct2})):
\begin{equation}
C^{IV}_N = \left(R_N C^{II}_N D_N\right)^T =  D_N C^{III}_N R^T_N\,. \label{eq:dct4_via_dct2_inv}
\end{equation}
By plugging this expression in (\ref{eq:dct2_basic_split}), we arrive at the following alternative decimation scheme:
\begin{equation}
C_N^{II} = P_N
\left(
\begin{array}{cc}
\!C^{II}_{N/2} & 0 \\
0 & D_{N/2} C^{III}_{N/2} R^T_{N/2} J_{N/2}
\end{array}
\right)
B_N. \label{eq:our_thing}
\end{equation}

Since only the order of operations has changed, the complexity of this decimation scheme and one used in C.\,W.\,Kok's algorithm (\ref{eq:kok}) must be exactly the same.
At the same time, as shown in Fig.\,\ref{fig_2}, this modified factorizations moves all the factors associated with matrix $D_{N/2}$ to the last stage. This means, that if it is sufficient to compute a scaled version of the transform, such multiplications can be avoided. Proposed factorization, therefore, is well suitable for implementation of scaled transforms.

Hereafter, we will say that DCT factorization is {\em scaled\/}, if it can be presented as
\begin{equation}
C^{II}_N = \Pi_N \Delta_N \tilde{C}^{II}_N \,, \label{eq:scaled_dct2}
\end{equation}
where $\Pi_N $ is a {\em reordering matrix\/}, and $\Delta_N $ is a diagonal {\em matrix of scale factors\/}, and
$\tilde{C}^{II}_N$ is a {\em matrix of the scaled transform\/}.

By using such representation, we can rewrite~(\ref{eq:our_thing}) as
\begin{equation*}
\Pi_N \Delta_N \tilde{C}^{II}_N
= \left(
\begin{array}{cc}
\!\Pi_{N/2} \Delta_{N/2} \tilde{C}^{II}_{N/2}\!\! & 0 \\
0 & \!\!D_{N/2} C^{III}_{N/2} R^T_{N/2} J_{N/2}\!
\end{array}
\right) B_N\,, \label{eq:scaled_dct2_decomposition}
\end{equation*}
implying, that scaled part of the transform can be computed recursively as follows
\begin{equation}
\tilde{C}_N^{II} =
\left(
\begin{array}{cc}
\tilde{C}^{II}_{N/2} & 0 \\
0 & C^{III}_{N/2} R^T_{N/2} J_{N/2}
\end{array}
\right)
B_N. \label{eq:scaled_rec}
\end{equation}
The associated reordering and scaling matrices can also be computed recursively by using
\begin{equation}
\Pi_N = P_N \left(
\begin{array}{cc}
\Pi_{N/2}\!\! & 0 \\
0 & \!\!I_{N/2}
\end{array}
\right)\,, ~
\Delta_N = \left(
\begin{array}{cc}
\Delta_{N/2}\!\! & 0 \\
0 & \!\!D_{N/2}
\end{array}
\right).  \label{eq:Delta_rec}
\end{equation}

In order to compute the remaining DCT-III block in (\ref{eq:scaled_rec}), we can either pick some existing (non-scaled) factorization, or reuse our scaled design (\ref{eq:scaled_rec}-\ref{eq:Delta_rec}) followed by conversion to full (non-scaled) transform
\begin{equation}
C^{III}_N = \left(\Pi_N \Delta_N \tilde{C}^{II}_N\right)^T = \tilde{C}^{III}_N \Delta_N \Pi^T_N  \,. \label{eq:DCTIII}
\end{equation}


%


\subsection{Complexity Analysis} \label{sec:IV_1}
As already noticed, the complexity of computing DCT-II by using our factorization (\ref{eq:our_thing}) is identical to one of C.W.Kok's algorithm (\ref{eq:mu_kok}). However, when only a scaled transform~(\ref{eq:scaled_rec}) needs to be computed, some operations can be saved. Based on Fig.\,\ref{fig_2}, we can establish the following relations:
\begin{eqnarray*}
\tilde{\mu}(N)    & = & \tilde{\mu}(N/2) + \mu(N/2)\,, \\
\tilde{\alpha}(N) & = & \tilde{\alpha}(N/2) + \alpha(N/2) + \tfrac{3}{2} N -1\,, \\
\tilde{\sigma}(N) & = & \tilde{\sigma}(N/2) + \sigma(N/2) + 1\,,
\end{eqnarray*}
where $\tilde{\mu}$, $\tilde{\alpha}$, and $\tilde{\sigma}$ denote the number of multiplications, additions, and shift operations correspondingly needed for computing scaled transforms $\tilde{C}^{II}$, and where $\mu$, $\alpha$, and $\sigma$ represent numbers of operations needed for computing lower non-scaled blocks $C^{III}$.
By applying this decomposition recursively $m$-times, we arrive at the following result.
\begin{proposition}
The numbers of arithmetic operations ($\mu,\alpha,\sigma$) needed for computing of \mbox{scaled DCT-II\/} of length $N=q\,2^m$ using factorization (\ref{eq:scaled_rec}) satisfy:
\begin{eqnarray}
\tilde{\mu}(N)    & = & \tilde{\mu}(q) + (2^{m}-1) \mu(q) + \left(\tfrac{m}{2}-1+2^{-m}\right) N \,, \label{eq:mu_q2m}\\
\tilde{\alpha}(N) & = & \tilde{\alpha}(q) + (2^{m}-1) \alpha(q) + \tfrac{3\,m}{2} N - 2^m +1\,, \nonumber \\ 
\tilde{\sigma}(N) & = & \tilde{\sigma}(q) + (2^m-1)\sigma(q) + 2^m-1\,. \nonumber 
\end{eqnarray}
\end{proposition}

By comparing (\ref{eq:mu_q2m}) with the number of multiplications required in C.\,W.\,Kok's algorithm (\ref{eq:mu_kok}), we can conclude that the use of our proposed scaled factorization saves at least
\begin{equation*}
\mu(N) - \tilde{\mu}(N) = \mu(q) - \tilde{\mu}(q) - \left(1-2^{-m}\right) N \geqslant \left(1-2^{-m}\right) N
\end{equation*}
multiplications. When number of iterations $m$ is large, it can be further observed that
\begin{equation*}
\mu(N) - \tilde{\mu}(N) \rightarrow N
\end{equation*}
approaching the well known upper bound for multiplicative complexity reduction realizable by scaled transforms~\cite{FeigWinograd}.

\begin{figure}[!t]
\centering
\resizebox{3.5in}{!}{\includegraphics{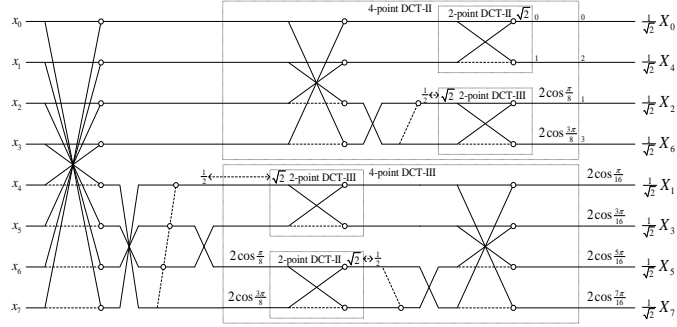}}
\caption{DCT-II of lengths $N=2^m$ ($m=1,2,3$) factorized by our algorithm. Dashed arrows show factors that can be merged.}
\label{fig_3}
\end{figure}

\begin{figure}[!t]
\centering
\resizebox{3.5in}{!}{\includegraphics{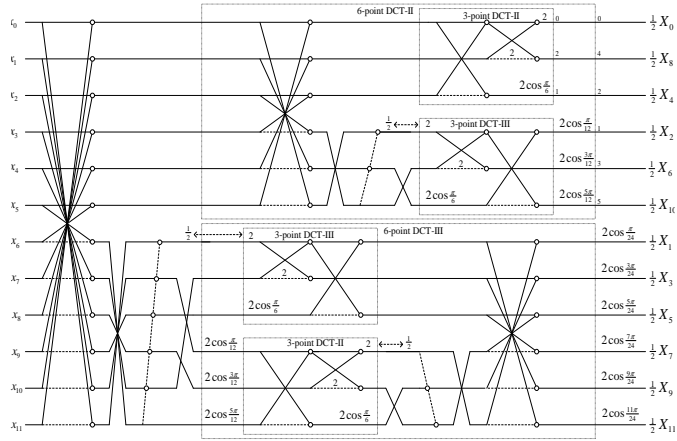}}
\caption{DCT-II of lengths $N\!=3\cdot 2^m$ ($m\!=\!1,2,3$) factorized by our algorithm. Dashed arrows show canceling factors.}
\label{fig_4}
\end{figure}

\subsection{Construction Examples} \label{sec:IV_2}
We note that in many practical situations, the multiplications by factors $1/2$ in our scheme can be avoided. Below, we provide two examples showing how this can be accomplished.

\subsubsection{Scaled DCT of lengths $N=2^m$} \label{sec:IV_2_A}
We scale the matrix of 2-point DCT-II as follows
\begin{equation*}
C^{II}_2 = \left(
\begin{array}{cc}
1 & 1 \\
\tfrac{1}{\sqrt{2}} & -\frac{1}{\sqrt{2}}
\end{array}
\right) =
\tfrac{1}{\sqrt{2}}
\left(
\begin{array}{cc}
\sqrt{2} & \sqrt{2} \\
1 & -1
\end{array}
\right).
\end{equation*}
This moves factors $\sqrt{2}$ in DC paths, allowing them to be subsequently merged with factors $1/2$ in our algorithm. We~show the resulting flowgraphs in Fig.~\ref{fig_3}.

Simple calculations show the number of operations in such scaled factorizations satisfy
\begin{eqnarray*}
\tilde{\mu}(2^m)  & = & m\,2^{m-1} - 2^m + 1 \,, \\
\tilde{\alpha}(2^m) & = & 3\,m\,2^{m-1} - 2^m +1\,, \\
\tilde{\sigma}(2^{m}) & = & 0\,.
\end{eqnarray*}
For example, when $N=8$ (largest size shown in Fig.~\ref{fig_3}) our algorithm produces factorization with just $5$ multiplications and 29 additions. This matches the performance of the well-known scaled DCT factorization of Y.\,Arai, T.\,Agui, and M.\,Nakajima~\cite{AAN}.


%

\subsubsection{Scaled DCT-II of lengths $N= 3\, 2^m$} \label{sec:IV_2_B}
We scale the matrix of 3-point DCT-II as follows
\begin{equation*}
C^{II}_3 = \!\left(
\begin{array}{ccc}
1 & 1 & 1 \\
\!\!\!\cos \frac{\pi}{6}\!\! & 0 & \!\!\!\!-\cos \frac{\pi}{6}\!\!\! \\
\frac{1}{2} & -1 & \frac{1}{2}
\end{array}
\right)\! =
\frac{1}{2}
\!\left(
\begin{array}{ccc}
2 & 2 & 2 \\
\!\!\!2\cos \frac{\pi}{6}\!\! & 0 & \!\!\!\!-2\cos \frac{\pi}{6}\!\!\! \\
1 & -2 & 1
\end{array}
\right)\!.
\end{equation*}
This brings factor $2$ to the DC path, leading to cancelation of factors $1/2$ in our algorithm. The resulting flowgraphs are shown in Fig.~\ref{fig_4}.

%
%
%
%
%
%

It can be readily verified that the numbers of operations in such scaled factorizations are
\begin{eqnarray*}
\tilde{\mu}\left(3\,2^m\right) & = & 3\,m\,2^{m-1} - 2^{m+1} + 2 \,, \\
\tilde{\alpha}\left(3\,2^m\right) & = & 9\,m\, 2^{m-1}+ 3\,2^m + 1\,, \\
\tilde{\sigma}\left(3\,2^m\right) & = & 2^m\,.
\end{eqnarray*}
For example, a scaled transform of length $N=6$ shown in Fig.~\ref{fig_4} uses only $1$ multiplication, $16$ additions, $2$ shifts.

\section{Comparison with the Prime Factor Algorithm-based Implementations} \label{sec:V}

It is known that DCT-II of length $N = p \, q$, where $p$ and $q$ are relatively prime, can be computed as a cascade of $p$ transforms of length $q$ followed by $q$ transforms of length $p$~\cite{YangNarasimha,BGLee89,RaoYip}. Such a  decomposition is commonly called a {\em prime factor algorithm\/} (PFA). When one of the prime factors, for example $p$, is dyadic, we arrive at lengths $N=q\,2^m$,
implying that PFA is an alternative technique for computing such transforms.
Hence, we are interested in comparison of PFA vs. C.\,W.\,Kok's algorithm.

We~report the following result.
\begin{theorem}
Multiplicative complexity of DCT-II of length $N=q \, 2^m$ constructed by using C.\,W.\,Kok's algorithm matches one theoretically achievable by using prime-factor DCT-II factorization, iff $m\leqslant 2$.
\end{theorem}
\begin{IEEEproof}
Based on PFA structure, the number of multiplications needed to implement transform of length $N=q \, 2^m$ satisfies (cf.~\cite{YangNarasimha,FeigLinzer})
$\mu
(N) = 2^m \, \mu(q) + q\, \mu(2^m)$. 
Furthermore, from complexity study of dyadic-length transforms \cite{Duhamel,HeidemanBurrus,FeigWinograd} we know that
$\mu\left(2^m\right) \geqslant 2^{m+1} -m -2$. 
Combining these formulae, we obtain
\begin{equation*}
\mu
(N) \geqslant 2^m \mu(q) + q \left( 2^{m+1} -m -2 \right). \label{eq:mu_pfa}
\end{equation*}
By comparing this result with complexity estimate for C.W.Kok's algorithm (\ref{eq:mu_kok}):
\begin{equation*}
\mu
(N) = 2^m \mu(q) +  q \,2^m \, \tfrac{m}{2}  \,,
\end{equation*}
we arrive at the statement of the theorem.
\end{IEEEproof}


We now turn our attention to complexity comparison for scaled transforms.
\begin{proposition}
Multiplicative complexity of PFA-based scaled DCT-II of length \mbox{$N = q\,2^m$} satisfies:
\begin{equation}
\tilde{\mu}(N) \leqslant  2^m \, \tilde{\mu}(q) + %
\tfrac{5}{2} N - q \tfrac{m\,(m+3) + 5}{2} - 2^{m-1} + \tfrac{1}{2}\,. \label{eq:pfa_mu_scaled}
\end{equation}
\end{proposition}
\begin{IEEEproof}
We use scaled PFA construction of Feig and Linzer~\cite{FeigLinzer}, which yields:
$\tilde{\mu}
(N) \leqslant 2^m \, \tilde{\mu}(q) + q\, \tilde{\mu}(2^m) + \tfrac{1}{2}\left(N - 2^m - q + 1\right)$. 
We then apply Feig-Winograd algorithm for computing scaled DCT of dyadic lengths~\cite{FeigWinograd}, for which:
$\tilde{\mu}\left(2^m\right) = 2^{m+1} - \tfrac{m\,(m+3)}{2} -2$. 
\end{IEEEproof}

\begin{table}[!t]
\caption{Component short-length DCT-II~\cite{Heideman,FeigLinzer,FeigWinograd,ChivukulaReznik,DCC09}}
\label{table_example}
\centering
\begin{tabular}{|c|ccc|ccc|}
\hline
$N$ & \multicolumn{3}{|c|}{DCT} & \multicolumn{3}{|c|}{Scaled DCT} \\
\cline{2-7}
     & $\mu$ & $\alpha$ & $\sigma$ &  $\tilde{\mu}$ & $\tilde{\alpha}$ & $\tilde{\sigma}$ \\
\hline
$3$  & $1$  & $4$  & $1$ & $0$ & $4$ & $1$   \\
$5$  & $4$  & $13$ & $1$ & $2$ & $13$ & $1$  \\
$15$ & $14$ & $70$ & $4$ & $10$ & $67$ & $8$ \\
\hline
$2$  & $1$  & $2$ & $0$ & $0$  & $2$ & $0$   \\
$4$  & $4$  & $9$ & $0$ & $1$  & $9$ & $0$   \\
$8$  & $11$  & $29$ & $0$ & $5$ & $29$ & $0$ \\
$16$ & $26$ & $81$ & $0$ & $16$ & $81$ & $0$ \\
\hline
\end{tabular}
\end{table}

\begin{table}[!t]
\caption{Complexity of Scaled DCT-II Factorizations of Lengths $N=q2^m$}
\label{table_example}
\centering
\begin{tabular}{|ccc|ccc|ccc|}
\hline
$q$ & m & $N$ & \multicolumn{3}{|c|}{Proposed algorithm} & \multicolumn{3}{|c|}{Feig and Linzer~\cite{FeigLinzer}} \\
\cline{4-9}
    &   &     & $\tilde{\mu}$ & $\tilde{\alpha}$ & $\tilde{\sigma}$ &  $\tilde{\mu}$ & $\tilde{\alpha}$ & $\tilde{\sigma}$ \\
\hline
3 & 1 & $6$  & $1$ & $16$ & $2$ & $1$ & $16$ & $2$ \\
  & 2 & $12$ & $6$ & $49$ & $4$ & $6$ & $49$ & $4$ \\
  & 3 & $24$ & $22$ & $133$ & $8$ & $22$ & $133$ & $8$ \\
  & 4 & $48$ & $66$ & $337$ & $16$ & $\mathbf{63}$ & $337$ & $16$ \\
\hline
5 & 1 & $10$ & $6$ & $40$ & $3$ & $6$ & $40$ & $2$ \\
  & 2 & $20$ & $19$ & $109$ & $7$ & $19$ & $109$ & $4$ \\
  & 3 & $40$ & $55$ & $277$ & $15$ & $55$ & $277$ & $8$ \\
  & 4 & $80$ & $147$ & $673$ & $31$ & $\mathbf{142}$ & $673$ & $16$ \\
\hline
15 & 1 & $30$ & $\mathbf{24}$ & $181$ & $13$ & $27$ & $178$ & $16$ \\
   & 2 & $60$ & $\mathbf{67}$ & $454$ & $23$ & $76$ & $445$ & $32$ \\
   & 3 & $120$ & $\mathbf{183}$ & $1090$ & $43$ & $204$ & $1069$ & $64$ \\
   & 4 & $240$ & $\mathbf{475}$ & $2542$ & $83$ & $505$ & $2497$ & $128$ \\
\hline
\end{tabular}
\end{table}

We note, that in order to compare the obtained expression (\ref{eq:pfa_mu_scaled}) with one corresponding to our scaled version of C.W.Kok's algorithm (\ref{eq:mu_q2m}):
\begin{equation*}
\tilde{\mu}(N) = (2^{m}-1) \mu(q) + \tilde{\mu}(q) + \left(\tfrac{m}{2}-1+2^{-m}\right) N \,.
\end{equation*}
we need to know complexities of both scaled and non-scaled transforms of length $q$. For this purpose, we will use several short-length DCT-II modules with complexity numbers shown in Table~1. Such odd-length transforms can be found in~\cite{Heideman} ($N=3$), \cite{ChivukulaReznik} ($N=5$), and~\cite{FeigLinzer,DCC09} ($N=15$). Listed complexity numbers for dyadic-length transforms are from~\cite{FeigWinograd,FeigLinzer}.

In Table~2 we provide comparison of the resulting transforms of composite lengths.
Bold font is used to highlight best complexity numbers. It can be observed, that for $q=3,5$ our proposed algorithm shows identical complexity to Feig-Linzer scaled PFA implementations when $m \leqslant 3$. It becomes more complex for higher $m$. For $q=15$ and $m \leqslant 4$ it is shown that our proposed algorithm is more efficient (in multiplicative complexity sense) than scaled PFA implementations.

\begin{figure*}[!t]
\centering
\resizebox{4in}{!}{\includegraphics{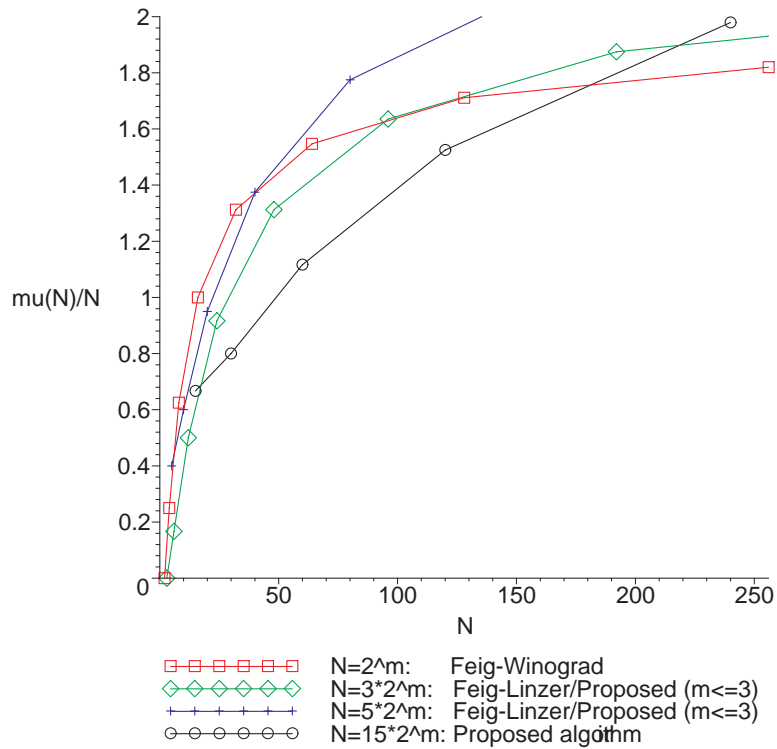}}
\caption{Normalized multiplicative complexity $\tilde{\mu}(N)/N$ of scaled DCT factorizations of lengths $N=[2^m, 3\, 2^m, 5\, 2^m, 15\, 2^m]$.}
\label{fig_5}
\end{figure*}

\section{On Normalized Multiplicative Complexity of Scaled Transforms} \label{sec:VI}
We complement our presentation by providing plots of normalized multiplicative complexity $\tilde{\mu}(N)/N$ of scaled DCT of lengths $N=[2^m, 3\, 2^m, 5\, 2^m, 15\, 2^m]$. We present these plots in Fig.~\ref{fig_5}.
It can be observed, that among short-length transforms ($N \leqslant 128$), {\em scaled dyadic-length transforms are more complex than transforms with nearest composite lengths from sequences $N=3\, 2^m$ or $N=15\, 2^m$\/}. We believe that the use of such composite-length transforms can offer appreciable complexity savings in many practical applications.

\section{Conclusions} \label{sec:VII}
An alternative derivation and detailed complexity analysis of C.\,W.~Kok's algorithm for computing DCT of lengths lengths $N=q\, 2^m$ ($m,q \in \mathbb{N}$, $q$ is odd) is offered. It is shown that this algorithm has the same multiplicative complexity as theoretically achievable by the prime factor decomposition, when $m \leqslant 2$.
Additionally, a scaled DCT factorization based on C.\,W.~Kok's algorithm is proposed. It is shown, that for some lengths this scaled factorization achieves lower multiplicative complexity than one of known prime factor-based scaled transforms.


%

%

\ifCLASSOPTIONcaptionsoff
  \newpage
\fi


\begin{thebibliography}{1}

\bibitem{AhmedNatarajanRao}
N. Ahmed, T. Natarajan, and K. R. Rao, ``Discrete Cosine Transform,'' {\em IEEE Trans. Computers}, vol.~X, pp.~90--93, Jan.~1974.

\bibitem{RaoYip}
K.R.~Rao, and P.~Yip, {\em Discrete Cosine Transform: Algorithms, Advantages, Applications\/}. New York: Academic Press, 1990.

\bibitem{BritanakYipRao}
V.~Britanak, P.~Yip, and K.R.~Rao, {\em Discrete Cosine and Sine Transforms: General Properties, Fast Algorithms and Integer Approximations\/}. Academic Press, 2007.


\bibitem{Duhamel}
P. Duhamel, ``New $2^n$ DCT algorithms suitable for VLSI implementation,''
{\em Proc. IEEE Int. Conf. Acoust. Speech, Signal Processing\/},
Dallas, TX, Apr.~1987, pp, 1805--1808.

\bibitem{HeidemanBurrus}
M.T.~Heideman, and C.S.~Burrus, ``On the number of multiplications necessary to compute a length-$2^n$ DFT,'' {\em IEEE Trans. Acoust., Speech, Signal Processing\/}, vol.~ASSP-34, no.~1, pp.~91--95, Feb.~1986.

\bibitem{FeigWinograd}
E.~Feig and S.~Winograd, ``On the multiplicative complexity of discrete cosine transforms (Corresp.),'' {\em IEEE Trans. Info. Theory\/}, vol.~IT-38, pp.~1387--1391, Jul.~1992.


\bibitem{ChenSmithFralick}
W.H.~Chen, C.H.~Smith, and S.C.~Fralick, ``A fast computational algorithm for the discrete cosine transform,'' {\em IEEE Trans. Comm.}, vol.~COM-25, pp.~121--123, Jan.~1983.

\bibitem{Wang83}
Z.~Wang, ``Reconsiderations of a fast computational algorithm for the discrete cosine transform,'' {\em IEEE Trans. Comm.}, vol.~COM-31, pp.~121--123, Jan.~1983.



\bibitem{BGLee}
B. G. Lee, ``A new algorithm for computing the discrete cosine transform,''
{\em IEEE Trans. Acoust., Speech, Signal Processing\/}, vol. ASSP-32,
pp. 1243–-1245, Dec. 1984.

\bibitem{LLM}
C. Loeffler, A. Ligtenberg, and G. S. Moschytz, ``Algorithm-architecture
mapping for custom DCT chips,'' {\em Proc. Int. Symp. Circuits
Syst., Helsinki, Finland\/}, June 1988, pp. 1953--1956.

\bibitem{AAN}
Y. Arai, T.Agui and M. Nakajima, ``A Fast DCT-SQ Scheme for Images'', {\em Transactions of the IEICE\/}, vol. E71, no. 11, p. 1095, Nov.~1988.

\bibitem{Feig}
E.Feig, ``A fast scaled DCT algorithm,'' {\em Proc. SPIE Int. Soc. Opt. Eng.\/}, vol 1244, pp. 2--12, 1990.

\bibitem{FeigWinogradSP}
E. Feig, S. Winograd, ``Fast algorithms for the discrete cosine transform,'' {\em IEEE Trans. Signal Processing\/}, vol. 40, no. 9, pp. 2174--2193, 1992.

\bibitem{FeigLinzer}
E. Feig, and E. Linzer, ``Scaled DCT's on Input Sizes that Are Composite,'' {\em IEEE Trans. Signal Processing\/}, vol. 43, no. 1, pp. 43--50, 1995.


\bibitem{ChanHo}
S.C.~Chan and K.L.~Ho, ``Direct methods for computing discrete sinusoidal transforms,'' {\em Proc. IEE\/}, vol.~137, pt.~F, no.~6, pp.~433--442, Dec.~1990.

\bibitem{kok}
C.W.~Kok, ``Fast algorithm for computing discrete cosine transform,'' {\em IEEE Trans. Signal Processing\/}, vol.~45, no.~3, pp.~757--760, Mar.~1997.


\bibitem{Heideman}
M.T.~Heideman, ``Computation of an Odd-Length DCT from a Real-Valued DFT of the Same Length,'' {\em IEEE Trans. Signal Processing\/}, vol. 40, no. 1, pp. 54--61, Jan 1992.

\bibitem{ChanHo93}
S.C. Chan and K.L. Ho, ``Fast algorithm for computing the discrete
cosine transform,'' {\em IEEE Trans. Circuits Syst. II\/}, vol. 44, pp. 185–-190,
Mar. 1993.

\bibitem{YangNarasimha}
P.P.N. Yang and M.J. Narasimha, ``Prime Factor Decomposition of the Discrete Cosine Transform,'' {\em Proc. IEEE Int. Conf. Acoust., Speech, Signal Processing\/}, Tampa, FL, March 26-29, 1985, pp. 772-775.

\bibitem{BGLee89}
B. G. Lee, ``Input and output index mapping for a prime-factor decomposed
computation of discrete cosine transform,'' {\em IEEE Trans. Acoust.,
Speech, Signal Processing\/}, vol. 37, pp. 237–-244, Feb. 1989.

\bibitem{ChivukulaReznik}
R.~K.~Chivukula,~Y.~A.~Reznik, ''Efficient implementation of a class of MDCT/IMDCT filterbanks for speech and audio coding applications,'' {\em  Proc. IEEE Int. Conf. Acoust., Speech, Signal Processing\/}, LasVegas, NV, March-April~2008, pp.~213-216.

\bibitem{DCC09}
Y.~A.~Reznik, and R.~K.~Chivukula, ``Fast 15x15 Transform for Image and Video Coding Applications'', {\em Proc. Data Compression Conference\/}, Snowbird, UT, March 16-18, 2009,
p. 465.

\end{thebibliography}
\end{document}